\documentclass[useAMS,usegraphicx,usenatbib,usedcolumn,preprint]{aastex}
\usepackage{spr-astr-addons}
\usepackage{lscape}
\usepackage{latexsym}
\usepackage{graphics}
\usepackage{sidecap}
\usepackage{subfigure}
\usepackage{textcomp}
\usepackage{bbding}
\usepackage{epsf}
\usepackage{amsmath,amssymb}
\usepackage{sidecap}
\usepackage{multirow}
\usepackage{flafter}
\usepackage{url}
\usepackage{float}
\usepackage{color}
\makeatletter
\renewcommand\rmxaa{\ref@jnl{Rev. Mexicana Astron. Astrofis.}}
\renewcommand\aj{\ref@jnl{Astron.~J.}}
\renewcommand\apj{\ref@jnl{Astrophys.~J.}}
\renewcommand\apjl{\ref@jnl{Astrophys. J. Lett.}}
\renewcommand\apjs{\ref@jnl{Astrophys. J. Suppl. Ser.}}
\renewcommand\aap{\ref@jnl{Astron. Astrophys.}}
\renewcommand\aaps{\ref@jnl{Astron. Astrophys. Suppl. Ser.}}
\renewcommand\mnras{\ref@jnl{Mon. Not. R. Astron. Soc.}}
\renewcommand\pasp{\ref@jnl{Publ. Astron. Soc. Pac.}}
\renewcommand\pasa{\ref@jnl{Proc. Astron. Soc. Aust.}}%
\renewcommand\nar{\ref@jnl{New Astron. Rev.}}%
\makeatother

\newcommand{\HI}{H{\sc i}}
\newcommand{\HII}{[H{\sc ii}]}

\def\p0{\phantom{0}}

\def\cm3{cm$^{-3}$}

\def\12{$^{12}$CO}
\def\13co{$^{13}$CO}

\def\HII{H{\sc ii}}

\RequirePackage{color}

\newcommand\msun{$\mathrm{M}_{\odot}$}

\newcommand\Ha{H$\alpha$}

\DeclareRobustCommand{\ion}[2]{%
\relax\ifmmode
\ifx\testbx\f@series
{\mathbf{#1\,\mathsc{#2}}}\else
{\mathrm{#1\,\mathsc{#2}}}\fi
\else\textup{#1\,{\mdseries\textsc{#2}}}%
\fi}

\newcommand\wl[1]{$\lambda\;#1$~\AA}
\newcommand\wll[2]{$\lambda\lambda\;#1, #2$~\AA}
\newcommand\linrat{\mbox{[\ion{S}{ii}]$_\mathrm{Total}$:\Ha}}
\newcommand\RA[3]{\mbox{$#1^\mathrm{h}#2^\mathrm{m}#3^\mathrm{s}$}}

\newcommand\DEC[3]{\mbox{$#1^\circ#2^\prime#3^{\prime\prime}$}}

\newcommand\Deg[1]{\mbox{$#1^\circ$}}
\newcommand\Min[1]{\mbox{$#1^\prime$}}
\newcommand\Units[1]{$\mathrm{#1}$}

\newcommand\ATCA[2]{J00#1$-$37#2}
\newcommand\NS{N300-S}

\newcommand\nosig{$\ldots$}
\newcommand\Sim{$\sim$}

\slugcomment{Submitted to Astrophysics and Space Sciences, \today.}

\shorttitle{Optical Spectra of SNR Candidates in NGC\,300}
\shortauthors{W. C. Millar, et al.}

\begin{document}

\title{Optical Spectra of Supernova Remnant Candidates\\in the Sculptor Group Galaxy NGC\,300}

 \author{William C. Millar\altaffilmark{1}}
  \affil{Centre for Astronomy, James Cook University, Townsville, Queensland 4811, Australia}
   \email{wmillar@grcc.edu}
\and
 \author{Graeme L. White, Miroslav D. Filipovi\'c\altaffilmark{2}, Jeffrey L. Payne, Evan J. Crawford}
  \affil{University of Western Sydney, Locked Bag 1797, Penrith South DC, NSW 1797, Australia}
\and 
 \author{Thomas G. Pannuti and Wayne D. Staggs}
        \affil{Department of Earth and Space Sciences, Space Science Center, 235 Martindale Drive, Morehead State University, Morehead, KY 40351, USA}

\altaffiltext{1}{Grand Rapids Community College, 143 Bostwick N.E., Grand Rapids, MI, 49503, USA}
\altaffiltext{2}{Centre for Astronomy, James Cook University, Townsville, Queensland 4811, Australia}

\begin{abstract}

We present moderate-resolution ($< 5$\AA) long-slit optical spectra of 51 nebular objects in the nearby Sculptor Group galaxy NGC\,300 obtained with the 2.3~meter Advanced Technology Telescope at Siding Spring Observatory, Australia. Adopting the criterion of \linrat~$\ge0.4$ to confirm supernova remnants (SNRs) from optical spectra, we find that of 28 objects previously proposed as SNRs from optical observations, 22 meet this criterion with six showing \linrat\ of less than 0.4. Of 27 objects suggested as SNRs from radio data, four are associated with the 28 previously proposed SNRs. Of these four, three (included in the 22 above) meet the criterion. In all, 22 of the 51 nebular objects meet the \linrat\ criterion as SNRs while the nature of the remaining 29 objects remains undetermined by these observations. 
\end{abstract}


%========================================================================
\section{Introduction}

{
The spectral identification of supernova remnants (SNRs) was pioneered in a series of papers by \citet{1972ApJ...178L.105M, 1973ApJ...179...89M, 1973ApJ...180..725M, 1973ApJ...182..697M} where narrow-band optical interference filters, centered on \Ha\ and the [\ion{S}{ii}] (\wll{6717}{6731}) doublet, were used to differentiate between primordial hydrogen and heavy metal contaminated ejecta of a SNR. This technique depended on the strength of the [\ion{S}{ii}] lines in SNRs being about the same strength as the \Ha\ lines, probably due to shock fronts in the expanding SNR shell, which in \ion{H}{ii} regions would not exist. The [\ion{S}{ii}] lines should be at least an order of magnitude weaker than the \Ha\ line in \ion{H}{ii} regions as compared to SNRs \citep{1972ApJ...178L.105M}. The \Ha\ filters are often was not able to remove the [\ion{N}{ii}] (\wl{6584}) line which is adjacent to the \Ha\ line. In some SNRs this line can be a strong as the \Ha\ itself. An emission region was classified as an SNR if it contained a (non-thermal) radio source and the \Ha\ + [\ion{N}{ii}] to [\ion{S}{ii}] ratio was less than two \citep{1972ApJ...178L.105M}.}

{\citet{1978AA....63...63D} pointed out the possibility of confusion between possible SNRs and \HII\ regions as well as between SNRs and shells of ionized gas originating as a consequence of supersonic stellar winds. They presented arguments based on observations of SNRs and \HII\ regions within the Galaxy and within the Large Magellanic Cloud to show how SNRs can be identified within M33 when \linrat $\leq0.4$. \citet{1985ApJ...292...29F} found that [\ion{O}{i}] \wll{6300}{6364}, [\ion{O}{ii}] \wl{3727} and [\ion{O}{iii}] \wll{4959}{5007} are often all simultaneously strong in SNRs and this can be used to differentiate SNRs from \HII\ regions in cases where \linrat\ is borderline.}

As we are located in the dusty disk of the Milky Way, the study of SNRs is hampered by the extinction and reddening effects of the interstellar medium (ISM) which impairs our ability to see Galactic SNRs at wavelengths other than radio. However, observing SNRs in nearby galaxies -- particularly in face-on galaxies with high Galactic latitude -- reduces absorption by both the host galaxy and our own \citep{Matonick1997, Pannuti2000}. Surveys of SNRs in the Local Group galaxies and galaxies within some nearby clusters have resulted in samples that are free from biases. A list of over 450 optical SNRs found in external galaxies is given by \citet{2005A&A...435..437U}, \citet{Matonick1997} and \citet{2007AJ....133.1361P}.

Finding new SNRs is a task best carried out using multiple wavelength surveys (mainly X-ray, optical and radio) rather than a single wavelength survey \citep[see][]{Filipovic1998, Lacey2001, Payne2006, Filipovic2008}. Radio-continuum observations using one frequency cannot uniquely identify SNR candidates, clearly differentiate SNRs from other nebulous objects or contend with the confusion that arises due to the presence of background sources (namely distant AGNs). An example of the most recent work in multiple wavelength observations of extra-galactic SNRs (M\,33) is presented by \citet{2010ApJS..187..495L}.

\begin{figure*}[!h]
%  \centerline{\includegraphics{NGC300MapNew.eps}}
  \centerline{\includegraphics[angle=0,width=1.0\textwidth]{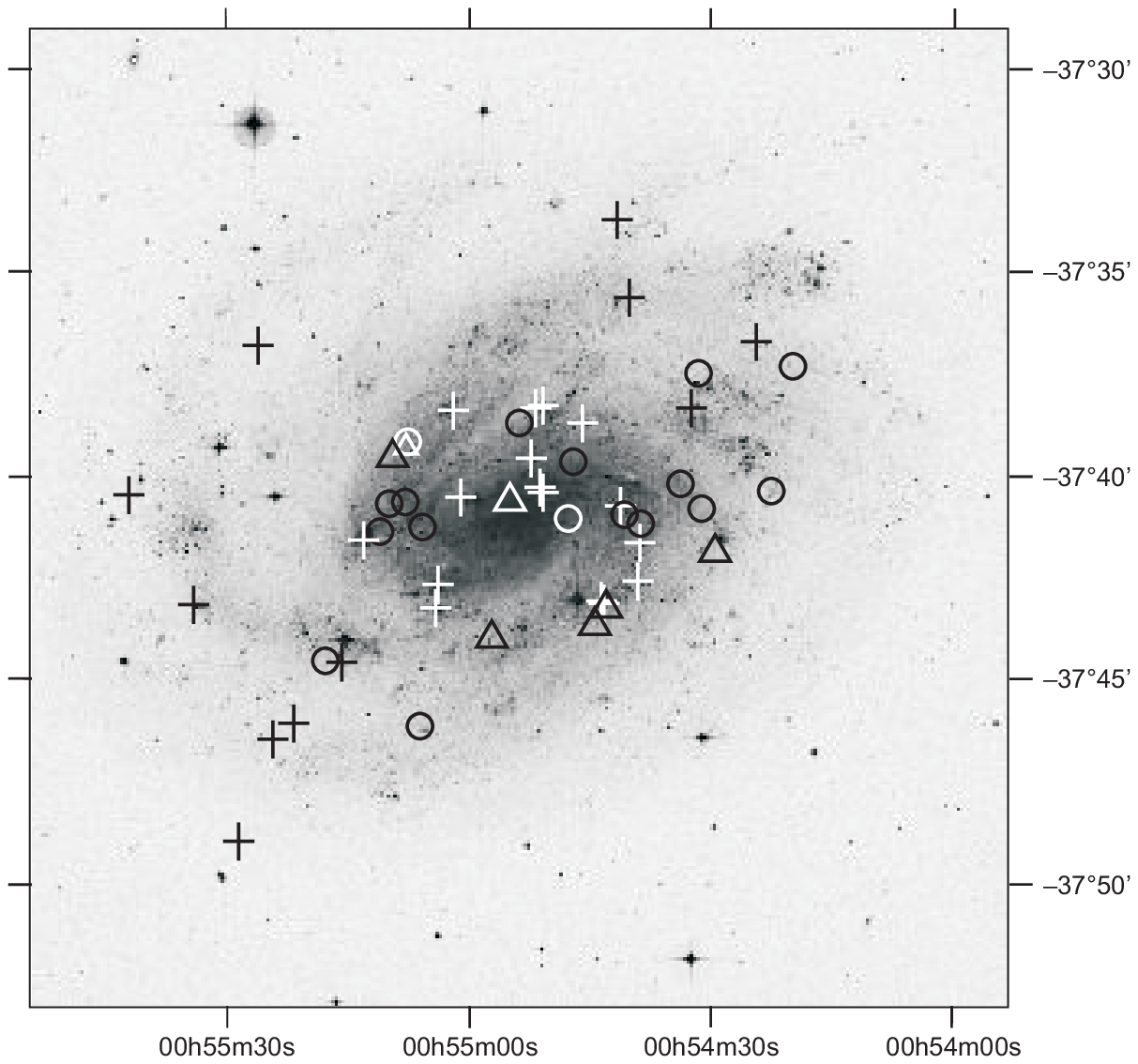}}
%  \plotone{Fig1.eps}
  \caption{A DSS image of NGC\,300 with the positions (in J2000.0 coordinates) indicated of the 51 SNRs and candidate SNRs considered by the present study. Radio sources (SNRs and SNR candidates only) from P04 are shown with crosses. Optical candidates with line ratios measured with long-slit spectra (from BL97) are shown as circles and optical candidates with line ratios measured by interference filters (BL97) are shown with triangles. Symbols are black or white only for increased contrast. (Southern sky DSS image, Royal Observatory Edinburgh, Anglo-Australian Observatory, California Institute of Technology.)}
%  \label{fig:TargLoc}
  \label{fig:1}
\end{figure*}

In this paper we present moderate-resolution ($<5$~\AA) long-slit optical spectra for 51 nebular objects in the nearby Sculptor Group galaxy NGC\,300. {\citet{1980AAS...40...67D} first published optical observations of SNRs in this galaxy. None of the 7 candidates published in that paper match any of the 51 candidates discussed here. The 51 candidates studied here were chosen from those published by \citet[hereafter BL97]{Blair1997} and \citet[hereafter P04]{Payne2004}.} Table~\ref{tab:1} provides a brief list of the characteristics of \objectname{NGC\,300} following \citet[hereafter P04]{Payne2004}; a more complete list is provided by \citet{Kim2004}. NGC\,300 is a type SA(s)d galaxy with an angular extent of 21.9\arcmin$\times$15.5\arcmin\ \citep[based on UV isophotes,][]{dePaz2007}. An image of NGC\,300 is presented in Figure~\ref{fig:1}: this galaxy has been classified as flocculent -- that is, its arms are poorly defined and it features many giant \HII\ regions which are evidence of many star formation episodes \citep[hereafter RP01]{Read2001b}. The similarity of angular size between NGC\,300 and other nearby spiral galaxies that have been studied (such as M\,33 and other members of the Sculptor Group, such as NGC\,7793) lead us to conclude that NGC\,300 is a typical, normal spiral galaxy (BL97). 

{We adopt the long accepted \linrat\ line flux density ratio to distinguish between a SNR and either a \HII\ region or a planetary nebula (PN). This criterion -- if the \linrat\ ratio is $\geq0.4$ the object is assumed to be a SNR, if $<0.2$ the object is more likely to be a \HII\ region or a PN -- } 
{has been used many times for galaxies in the Local Group and other nearby galaxies \citep[for example BL97;][]{Matonick1997, McNeil2006, Payne2007, Payne2008a, Payne2008b}. As described previously, the
physical processes thought to create this criterion are well understood (BL97 and references therein).}

Because of its low inclination angle (measured to be between \Deg{43} and \Deg{46}; \citet{Tully1988} and \citet{Puche1990}) and its high Galactic latitude (\Deg{-77.17}, Table~\ref{tab:1}), observations of NGC\,300 entail very low internal extinction \citep{Butler2004} and foreground reddening \citep[$E(B-V)=0.013$ mag,][]{Bland2005}. For studies of this galaxy, most authors have adopted distances of 2.0-2.1 Mpc \citep[BL97;][]{Freedman1992, Freedman2001} though recent distance measurements based on observations of Cepheid variables have favoured a nearer distance of 1.88~Mpc with an error of 3\% \citep{Bresolin2005, Gieren2005}. For the present paper, we have adopted a distance of 2.1~Mpc to be consistent with previous observations (BL97, P04). The corresponding linear scale is 10.2~\Units{pc\;arcsec^{-1}}.

The observations and the corresponding (optical and X-ray) data reduction are presented in Section~2. The results of this analysis are presented in Section\,3 including notes on individual objects. We confirm 22 objects as SNRs while the nature of the remaining 29 objects remains uncertain. Finally, in Section\,4, we summarize our main results.

%==================================================================================================

\section{Observations and Data Reduction}

\subsection{Optical Data}

The positions (J2000.0) of all 51 SNRs and SNR candidates observed are given in Table~\ref{tab:2} and are shown in Figure~\ref{fig:1}. Optical spectra of all the sources were obtained in August of 2006 using the 2.3~m ($f$/17.9) Advanced Technology Telescope at Siding Spring Observatory, Australia, and the Dual-Beam Spectrograph \citep[DBS,][]{Rodgers1988}. The slit width was 1~arcsec and a 4~arcmin decker was adopted. The wavelength domain used was 5400~\AA\ to 9000~\AA\ (from which we extracted data between 6300~\AA\ and 6800~\AA).

Most objects were observed twice, once with the DBS slit aligned in Declination (position angle, PA=$\Deg{0}$) and again with the slit aligned in Right Ascension (PA=$\Deg{90}$). A few objects were surrounded by interesting structure (in DSS2-Red) and observations were made with the slit positioned at the appropriate angle. Telescope pointing was confirmed by comparing the DBS slit camera image with pointing charts prepared from the DSS2-Red.

The observations reported here used only the red arm of the DBS. The grating used was the 316R (316 lines/mm) which is blazed at \Deg{6} \Min{48}. This grating gives a resolution of 4.1 \AA\ (170 \Units{km\;s^{-1}}). Used here at a Grating Angle of \Deg{3} \Min{49}, the 316R grating gives a central wavelength $\mathrm{\lambda_{cent}}$ of 7200~\AA. The DBS's SiTE CCD camera has dimensions (1752$\times$532 pixels, with pixels of $15\mu$m). All spectra were 600 second exposures returning a typical background (sky) level of \Sim30--60 counts per pixel and a center of spectra level of \Sim400--600 counts per pixel.

Data reduction and analysis was performed using the Image Reduction and Analysis Facility (IRAF) software package, with Starlink's Figaro cosmic ray cleaner, and Brent Miszalski's ``Planetary Nebula Extraction'' package for IRAF. All line flux densities were measured with the DEBLEND function of IRAF's SPLOT task. Data reduction included bias subtraction, flat-field correction and wavelengths calibration using standard NeAr arc-lamp lines. The star EG274 was the photometric standard \citep{Stone1983, Baldwin1984, Hamuy1994} for flux density calibration.

\subsection{X-ray Data}
 \label{xray}

To complement our optical spectroscopic observation of SNRs in NGC\,300, we also analyzed an archival X-ray observation made of this galaxy with the Chandra X-ray Observatory \citep{2002PASP..114....1W}. This observation made use of the High Resolution Camera (HRC-I) \citep{2000SPIE.4012...68M} which can attain an angular resolution of approximately 0.4~arcseconds. Data for this observation (ObsID 7072 -- centered at RA(J2000)=00$^{\rm h}$55$^{\rm m}$10$^{\rm s}$, DEC(J2000)=--37$^{\circ}$38\arcmin55\arcsec) was downloaded and reduced using standard tools available in the Chandra Interactive Analysis of Observations (CIAO) package \citep{2006SPIE.6270E..60F} Version 4.0.1. The CIAO tool ``acis\_process\_events'' was run to apply the latest calibration files: in addition, the dataset was filtered based on grade and status to create a new level=2 event file (that is, events were removed that did not have a good grade or had one or more of the STATUS bits set to 1). The good time intervals and a light curve was generated to search for background light flares during the observation. The effective exposure time of the final image after processing was 15.19~kiloseconds and the corresponding energy range is 0.3--10.0~keV. To detect sources in this field of view, we ran the tool ``wavdetect,'' which is a wavelet-based algorithm used for source detection \citep{2002ApJS..138..185F}: a total of 31 sources were detected to a limiting unabsorbed luminosity of approximately 10$^{37}$ ergs/sec (assuming a distance of 2.1~Mpc to NGC\,300, a column density of $N$$_H$=3.08$\times$10$^{20}$ cm$^{-2}$ and a Raymond-Smith thermal plasma emission model with a temperature $kT$=0.5~keV and solar abundance ratios). 

From these 31 sources, only two discrete X-ray sources were found to match the positions (to within 2\arcsec\ or less) of known SNRs in NGC\,300. These particular SNRs are the optically-identified SNRs \NS10 and \NS26. The associations between the X-ray sources and these optically-identified SNRs have been presented in previous works (P00, P04). We estimate the absorbed (unabsorbed) luminosities (over the energy range of 0.3--10.0~keV) for the HRC-detected X-ray counterparts to S10 and S26 to be 2.7$\times$10$^{37}$ (3.5$\times$10$^{37}$) and 1.1$\times$10$^{37}$ (1.5$\times$10$^{37}$), respectively. These luminosities were calculated using the tool PIMMS\footnote{http://cxc.harvard.edu/toolkit/pimms.jsp} Version~3.9i assuming a Galactic column density N$_{H}$ = 3.08$\times$10$^{20}$cm$^{-2}$ toward NGC\,300 and a thermal bremsstrahlung model with a temperature of kT=0.5~keV. Unfortunately, because of the short exposure time of the observation, the limiting unabsorbed luminosity ofapproximately 1$\times$10$^{37}$ergs\,sec$^{-1}$ is too high to detect X-ray emission from the large majority of SNRs associated with NGC\,300.

%==================================================================================================

\section{Analysis and Results}

As described previously, at optical wavelengths, SNRs are identified primarily by the flux density ratio of [\ion{S}{ii}]:\Ha. When this ratio is greater than 0.4, the nebula is considered to be as a SNR %\citep[BL97]{Matonick1997}
and the presence of some other optical spectral lines may lend support to the classification. Table~\ref{tab:3} is the collected multi wavelength observations for the 51 objects of Table~\ref{tab:2} as selected principally from P04 and BL97. {Tables~\ref{tab:3}, \ref{tab:4} and \ref{tab:5} are divided into three sections according to the results of our observations. The first is ``SNRs'' which are those objects for which our observations resulted in a \linrat $\geq0.4$. The second is those objects which did not meet this requirement and the third is those objects for which we obtained no spectrum.} In Table~\ref{tab:3}; Column~1 is the source name adopted by BL97 and Column~2 is the radio source name from P04, based on the J2000 position. Columns~3 and 4 give the radio data for the sources with Column~5 listing the nature of the source proposed by P04. Columns~6 through 9 show X-ray observations and Columns~10 through 12 show previous optical observations. There is no standard designation style for X-ray sources, so X-ray nomenclatures in Columns~6 through 9 are tied to individual papers. 

Column 3 gives the radio spectral index from Very Large Array (VLA) observations at 1465 and 4885~MHz as reported in P00. Column 4 shows the radio spectral index as reported in P04. These values are based on flux densities obtained at 1347, 1448, 2496 and 4860~MHz at the Australia Telescope Compact Array (ATCA, P04) or the VLA (P00). 

Column 5 gives the object type proposed by P04 based on their radio spectral index of Column~4. Sources are classified as candidate radio SNRs if the spectral index $\alpha$ \footnote{Where $\alpha$ is defined as $S\propto\nu^\alpha$} of the radio emission is in the range $-0.8 \leq \alpha \leq -0.2$ and if it is co-identified with an X-ray source. This range is based on a statistical average of the spectral indices of over 270 Galactic SNRs \citep{Trushkin1998}. P04 also classified radio SNRs taking in account their association with known optical SNR, OB association or \HII\ region.  Other object types proposed by P04 are ``snr'' -- radio SNR candidate; ``snr$\dagger$'' -- SNR based on spectral index only; ``\HII'' -- \HII\ region; ``hii'' -- possible \HII\ region; ``BKG'' -- background radio source or ``bkg'' -- possible background radio source.

All SNRs emit soft X-rays resulting from heated gas inside the expanding shock front \citep{Aller1991,Osterbrock2006}. The identification of potential SNRs at X-ray wavelengths is based on spectral fits to the observed emission using thermal bremsstrahlung models. X-ray emission can also occur from SNRs by virtue of an embedded pulsar or neutron star \citep[e.g.]{2003ApJ...594L.111G}. X-ray emission from SNR candidates is usually fitted to models (temperature and particle density) of these emission types to verify the candidacy.

Columns 6 through 9 show data reported by four papers of X-ray point sources in positional agreement with those sources observed in this paper. Data comes from \citet[hereafter RPS97]{Read1997}, \citet[hereafter RP01]{Read2001b}, P04 and \citet[hereafter C05]{Carpano2005}, respectively.

Data in Columns 6 and 7 were generated from observations made with the ROSAT X-ray observatory while data in Column~8 are from results obtained with observations made with the XMM-Newton observatory. In addition, we have further investigated putative associations between X-ray sources and SNRs through the analysis of an archival Chandra HRC-I observation of NGC\,300 (also see Sect~\ref{xray}).

Column 10 indicates if the source is visible as a nebulous object on the DSS2-Red survey. Column 11 gives the ratio of [\ion{S}{ii}] to \Ha\ line flux density as reported by BL97. The values in parenthesis are from Table~3A of BL97 and are based on interference filter images, otherwise they are from Table~4A of BL97 and are based on long-slit spectra.

For approximately half of the spectra collected the length along the decker of the \Ha\ line (\wl{6563}), and the [\ion{N}{ii}] (\wll{6548}{6583}) and [\ion{S}{ii}] (\wll{6716}{6732}) lines were essentially the same, indicating that the emission regions were of approximately the same physical size. Reduction of these spectra followed standard procedures.

For spectral lines where the \Ha, [\ion{N}{ii}] and [\ion{S}{ii}] lines differ in length (indicating a possible different physical size for the emitting regions) extraction was done so as to ensure that the line ratio was not dominated by extracted \Ha\ emission from the background.

Table~\ref{tab:4} gives the integrated line flux densities for all objects in Table~\ref{tab:3}. The first two columns are the designation of the SNR or SNR candidate in P04 (Column~1) or BL97 (Column~2). Also listed is the integrated \Ha\ line (6563~\AA) flux density (Column~3); the total, integrated [\ion{N}{ii}] line (6548~\AA\ + 6583~\AA) flux density (Column~4); the total, integrated [\ion{S}{ii}] line (6716~\AA\ + 6731~\AA) flux density (Column~5); the ratio of \linrat\ (Column~6); the total, integrated [\ion{O}{i}] line (6300~\AA\ + 6364~\AA) flux density (Column~7); and the \Ha\ diameter of the object (Column 8). 

\begin{figure*}[!h]
%  \centerline{\includegraphics[scale=0.5]{Fig2BW.eps}}
%  \centerline{\includegraphics[scale=0.5]{Fig2.eps}}
  \centerline{\includegraphics[angle=-90,width=1.0\textwidth]{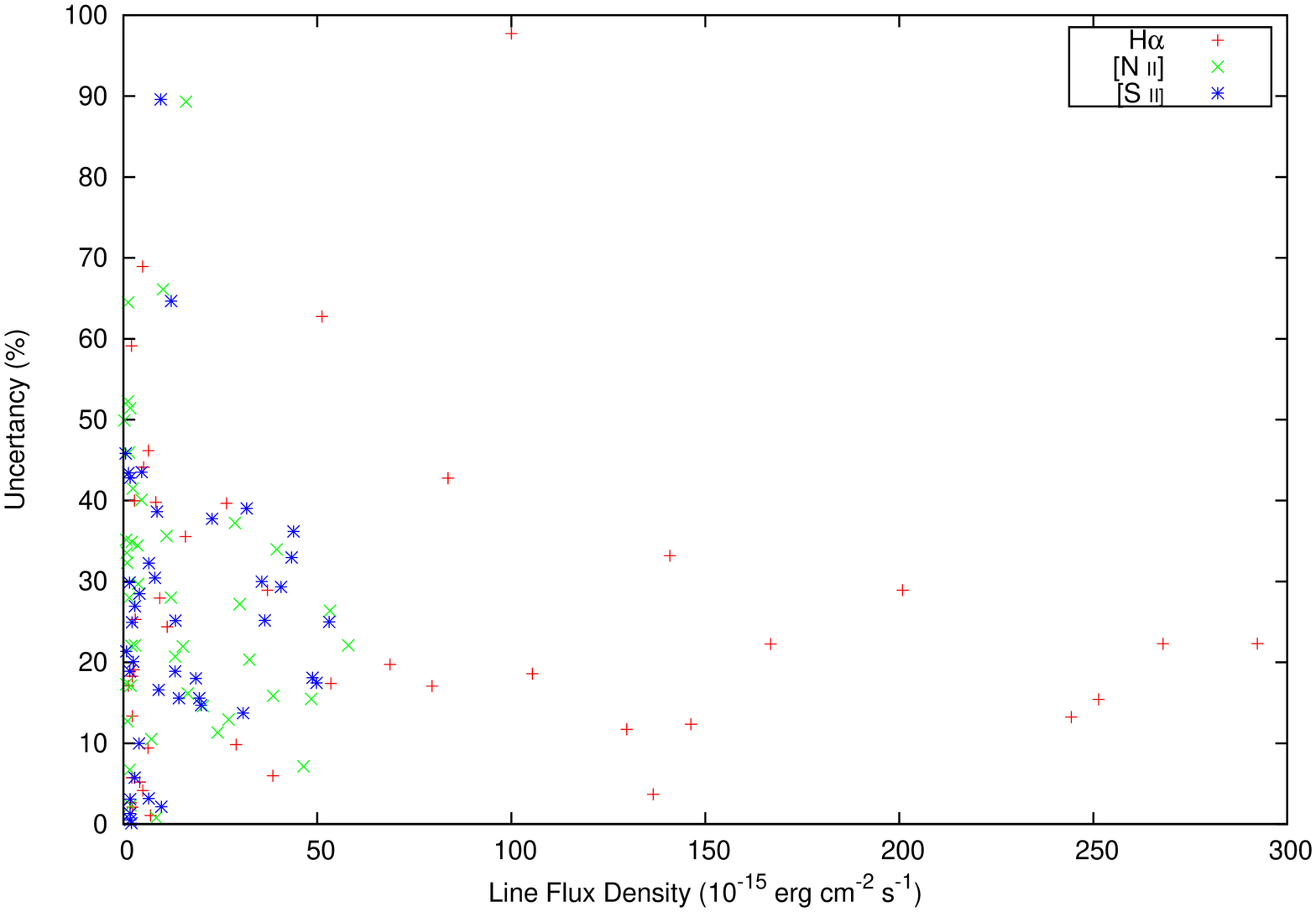}}
%  \centerline{\includegraphics[angle=-90,width=0.7\textwidth]{error-log.eps}}
%  \plottwo{Fig2BW.eps}{Fig2.eps}
  \caption{The standard errors (expressed as a percentage) in the means (SEMs) of the multiple flux density measurements in the individual spectral lines as a function of flux density of that line. The noise independent uncertainty is about 22\%. Details are given in the text.}
  \label{fig:2}
\end{figure*}

With a few exceptions, each object was observed at least twice and these multiple observations allow the direct computation of the uncertainties in flux densities of each line. Figure~\ref{fig:2} shows the standard errors (expressed as a percentage) in the means (SEMs) of the individual flux densities for each spectral line in Table~\ref{tab:4} as a function of flux density of that line. Here, the noise independent uncertainty (of about 22\%) and the noise dependent component are clearly delineated. The envelope of uncertainties in Figure~\ref{fig:2} is defined by: 
\begin{equation}
\rm{ \Delta L = \sqrt{0.15 \rm {L + 0.05 L}}}
\end{equation}
\noindent where $\rm {\Delta L}$ is the uncertainty in the flux density of each individual line and L is the flux density of that line in units of 10$^{-15}$ ergs cm$^{-2}$ s$^{-1}$. The formal uncertainties in the line flux density values of Columns 3, 4 and 5 of Table~\ref{tab:4} are defined by this relationship.

The uncertainty in the \linrat\ (Column 6 of Table~\ref{tab:4}) is governed by the uncertainties in the flux densities of the individual lines, and is about 8\% for ratios close to unity and about 20\% for ratios tending towards zero. For the smaller values of \linrat\ the uncertainty is made larger by the increased uncertainty in the weaker [\ion{S}{ii}] lines.

A comparison of the line ratios reported here with those of BL97 (Figure~\ref{fig:2}) shows consistency between our results and the results presented by BL97, but with a small (of order 10\%) bias (with BL97 greater than the current work) and an indication that this bias is induced by the size of the source as well as the selection of the type of extraction used (see above). As one would expect, the bias is also dependent on the strength of the lines, caused by the rising noise component in the line flux density for very weak lines.

\subsection{Measurement of the Supernova Remnant Diameter}

Column 7 of Table~\ref{tab:4} is our measured diameter of the SNR. We have attempted to estimate the linear diameter of the SNR or SNR candidate by fitting a Gaussian profile to the \Ha, [\ion{N}{ii}] and [\ion{S}{ii}] lines along the decker orthogonal to the wavelength directions. The image scale of the SiTE detector is 0.78~arcseconds per channel which corresponds to 7.7~pc per channel at the adopted distance of NGC\,300 (2.1~Mpc). Each spectral line was then deconvolved with the standard star for that night to give a diameter of the source in parsecs. The full-width half-maximum (FWHM) diameters are given in Column 7 of Table~\ref{tab:4}. 

For the optical candidates published in BL97, a comparison of our diameter measurements with those in BL97 does not show good correlation (r$^2=0.02$). This, we believe, is not the result of any problem with our data or that of BL97, rather that the data sets are limited by the seeing which is of approximately the same size as the object itself. Our measured diameters indicate an average size of 54$\pm$22~pc, and on this basis, we note that there are only three objects (\NS4, \NS11 and \NS24; where \NS4 and \NS24 are confirmed SNRs) that are worthy of being noted as large objects at 150~pc, 150~pc and 100~pc respectively. Sources of this size may be superbubbles rather than a single SNR.

In addition, we have looked for large \Ha\ diameters relative to smaller [\ion{N}{ii}], [\ion{S}{ii}] or [\ion{N}{ii}] + [\ion{S}{ii}] diameters, as indications of an embedded SNR in \HII\ regions. There is only one object that might stand a solid statistical analysis -- \NS6 -- where the \Ha\ is larger than the [\ion{S}{ii}] by a factor of 2.6 and the combined [\ion{N}{ii}] + [\ion{S}{ii}] is larger by 2.2. However, BL97 report the same diameter as us: therefore, either the \ion{S}{ii} diameter reported by BL97 or our \Ha\ diameter measurement is an overestimate. Further work at higher angular resolution is required to better determine the diameters of these objects.

\subsection{Overall Results}

The SNR candidates published in BL97 where originally found using \Ha\ and [\ion{S}{ii}] interference filters on the 2.5~m du~Pont Telescope at Las Campanas. Surveys of this type preferentially find objects with large \linrat\ ratios. As confirmation, moderate-resolution long-slit spectra were obtained by BL97 for 21 of their 28 candidates using the Modular Spectrograph and the line flux density measurements are presented in BL97 as Table~4A. We obtained satisfactory spectra for all 28 candidates and confirm 22 (78\%) as optical SNRs based on the \linrat.

\begin{figure*}[!h]
%  \centerline{\includegraphics{S429e.eps}}
  \centerline{\includegraphics{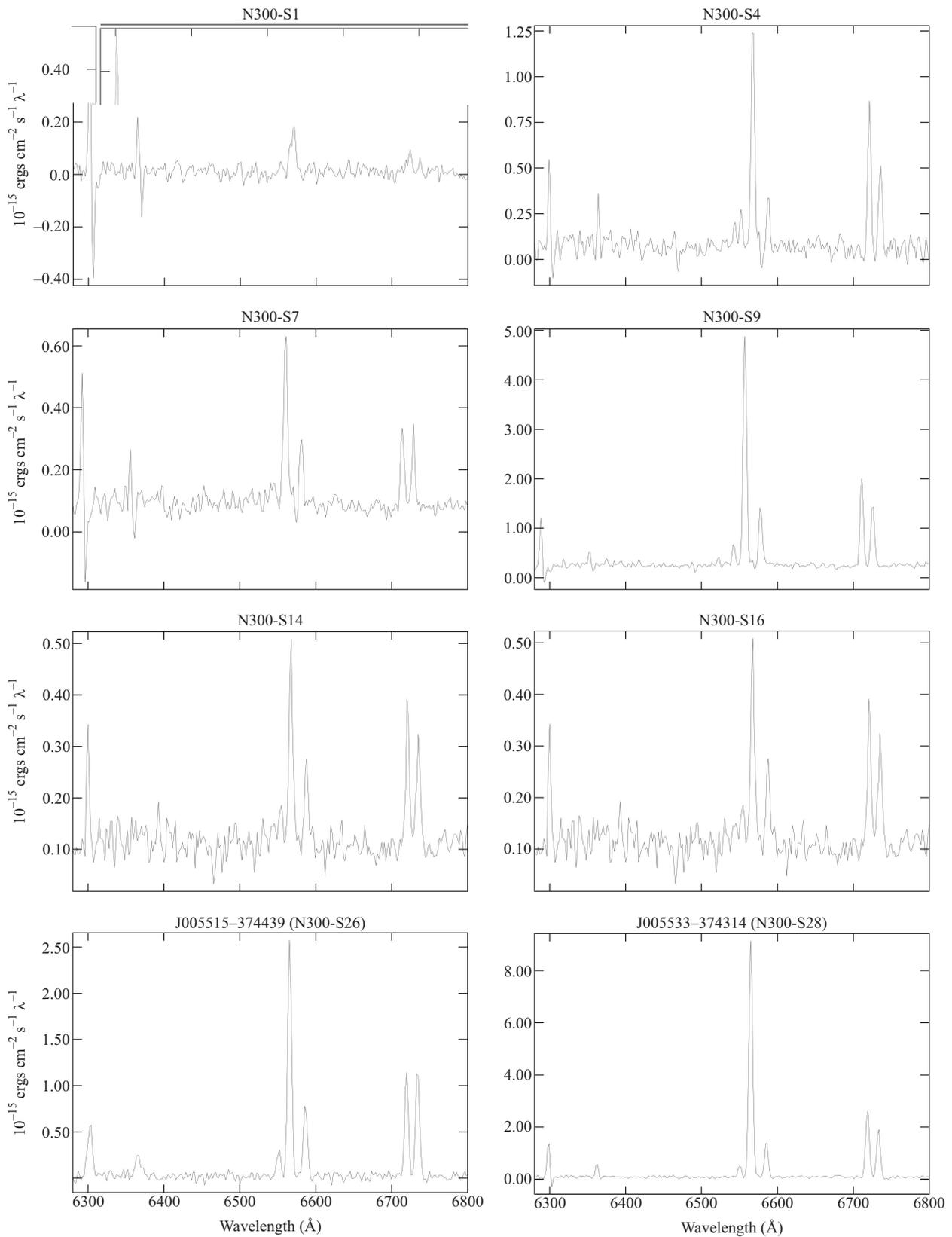}}
%  \epsscale{.80}
%  \plotone{Fig3.eps}
  \caption{Example spectra of objects meeting the \linrat~$>0.4$ criterium and are therefore labeled as SNRs in our results.}
%  \label{fig:S429}
  \label{fig:3}
\end{figure*}

\begin{figure*}[!h]
%  \centerline{\includegraphics{S429e.eps}}
  \centerline{\includegraphics{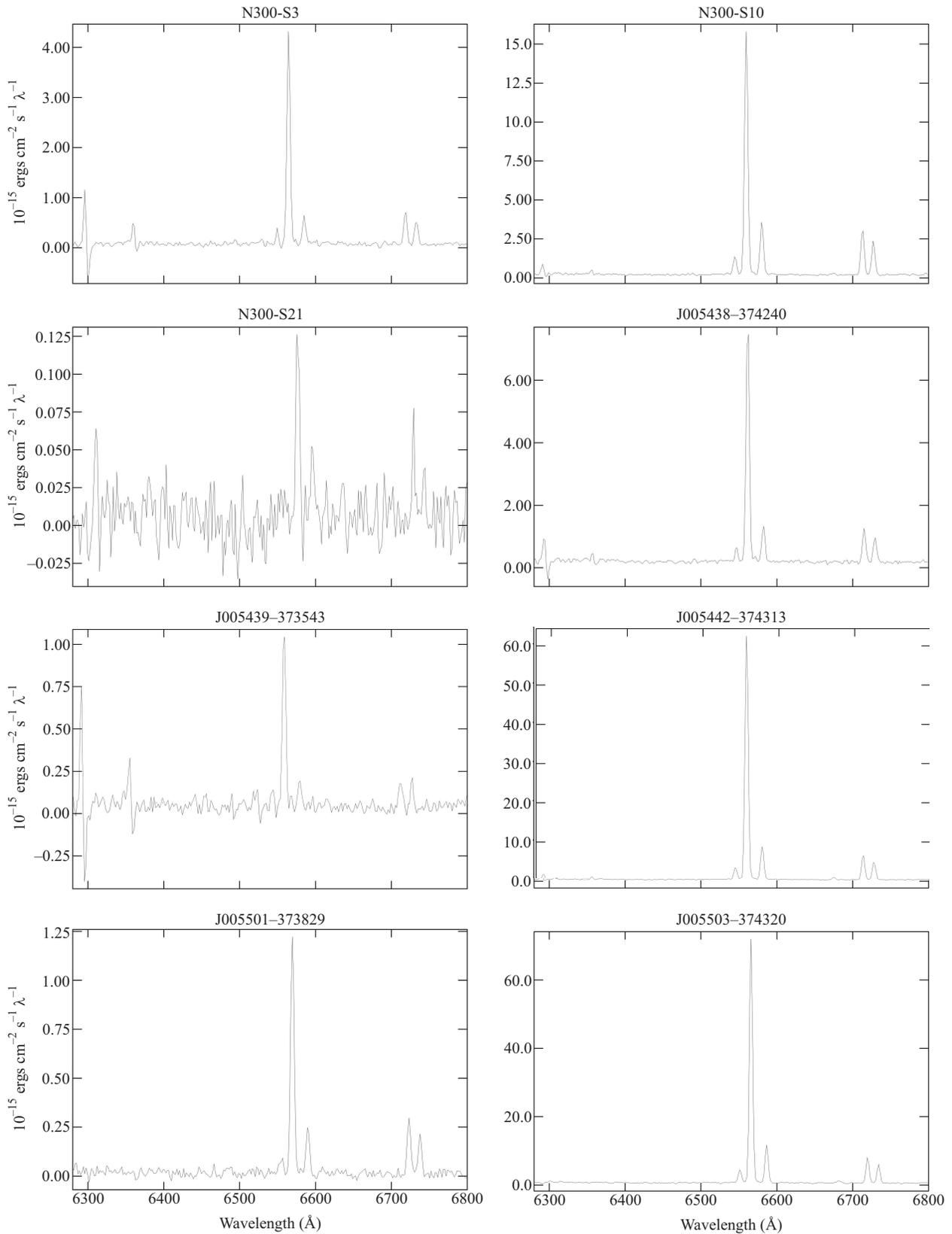}}
%  \epsscale{.80}
%  \plotone{Fig3.eps}
  \caption{Example spectra of objects which did not meet the \linrat~$>0.4$ criterium and are therefore labeled as ``other objects'' in our results.}
%  \label{fig:S429}
  \label{fig:4}
\end{figure*}

{Figure~\ref{fig:3} shows examples of spectra of objects which meet the \linrat $\geq0.4$ criterion and are therefore labeled as SNRs in our results. Figure~\ref{fig:4} shows examples of spectra for objects which do not meet the criteria.} A summary for our results is given in Table~\ref{tab:5}. On the basis of the above definition we confirm 22 objects as SNRs. Of the 27 radio objects suggested as SNRs by P04, we confirm only three (11\%). These three radio sources are positionally linked to three of the SNRs listed in BL97. 

\begin{figure*}[!h]
  \centerline{\includegraphics[angle=-90,width=\textwidth]{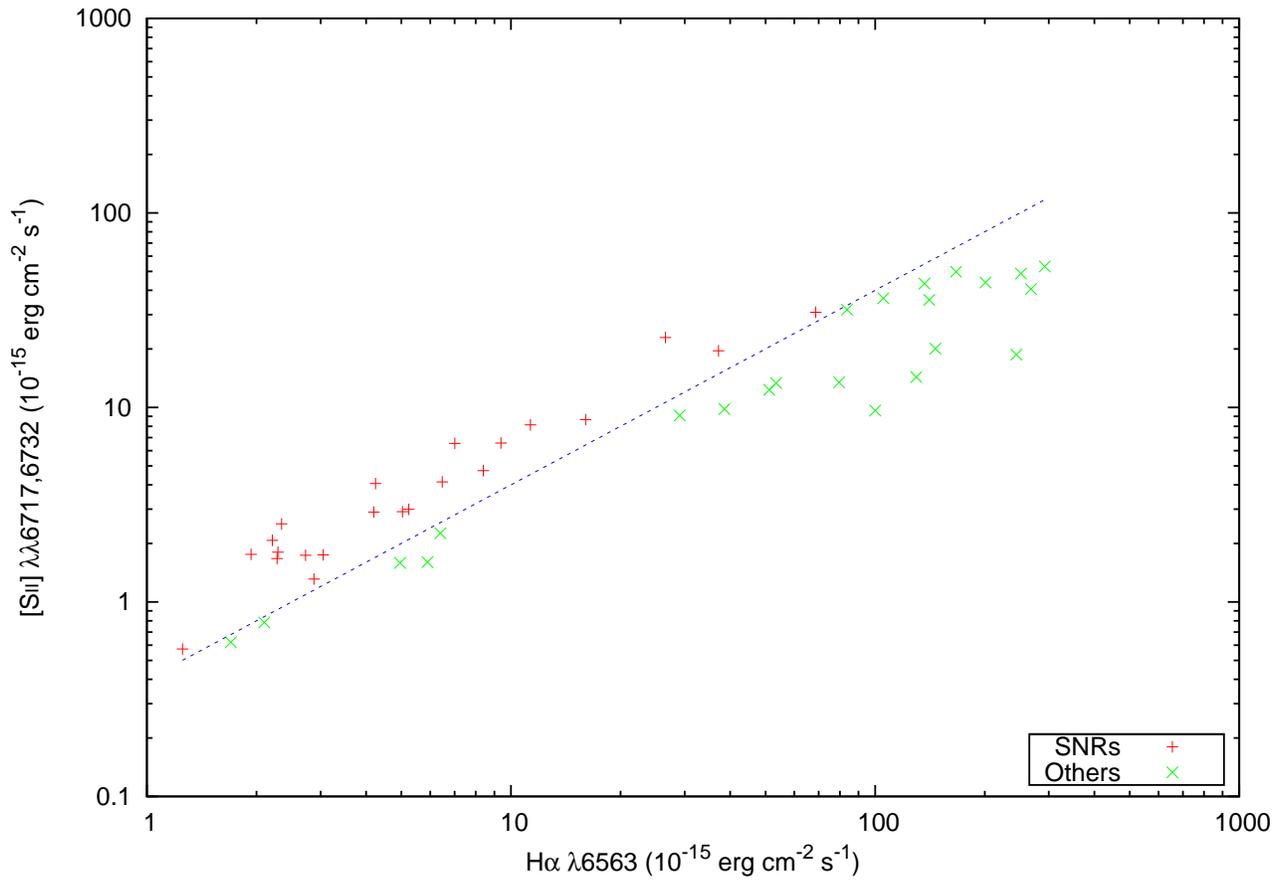}}
  \caption{A plot of \linrat\ values for the objects we label as SNRs and ``other objects.'' The scales were made logarithmic to make the data points more visible. The  dashed line  represents a \linrat\ of 0.4.}
  \label{fig:has2}
\end{figure*}

\begin{figure*}[!h]
  \centerline{\includegraphics[angle=-90,width=\textwidth]{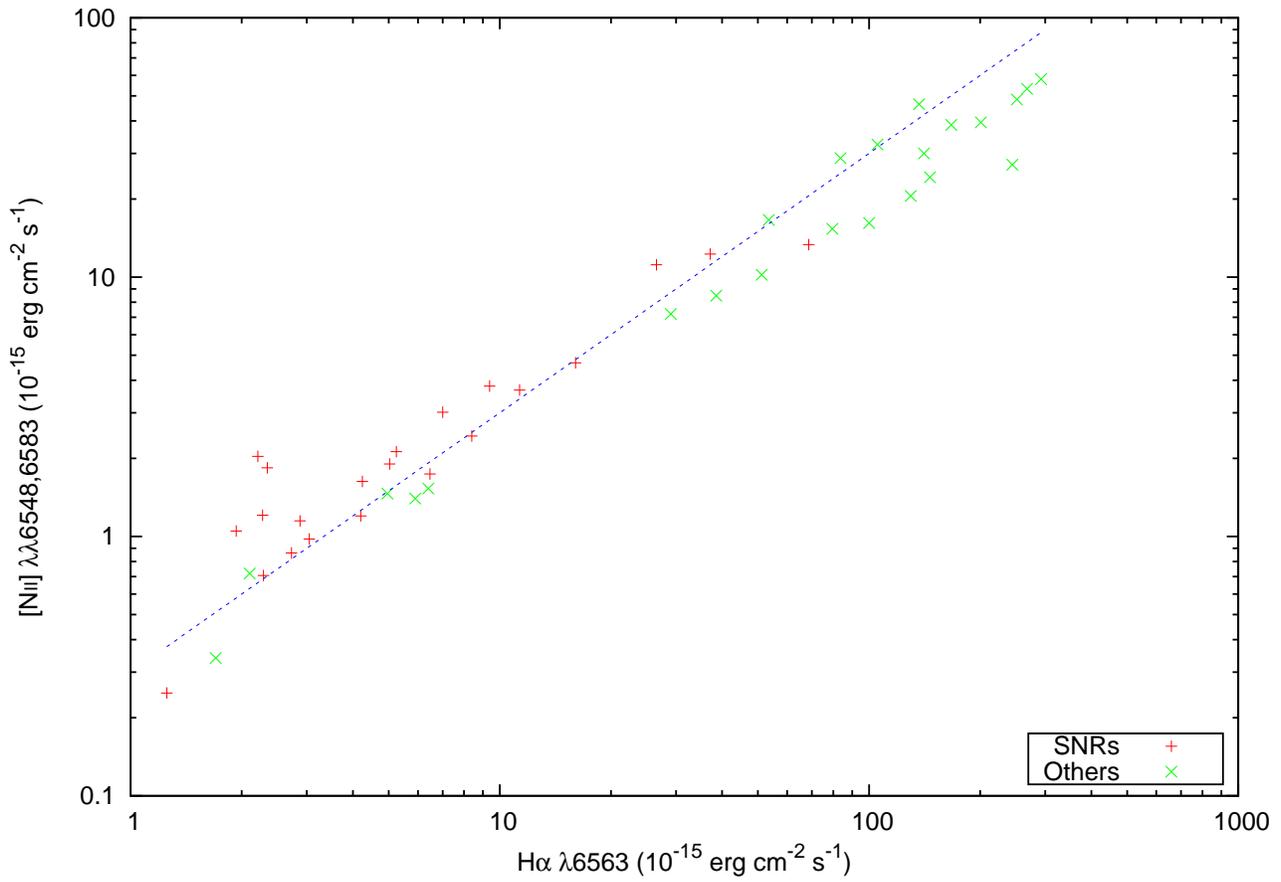}}
  \caption{A plot of [\ion{N}{ii}] vs. \Ha\ flux values for the objects we label as SNRs and ``other objects.'' The scales were made logarithmic to make the data points more visible. The plot shows a consistent nitrogen-hydrogen ratio of roughly 0.3 (indicated by the dashed line) across both groups of objects.}
  \label{fig:han2}
\end{figure*}

\begin{figure*}[!h]
  \centerline{\includegraphics[angle=-90,width=\textwidth]{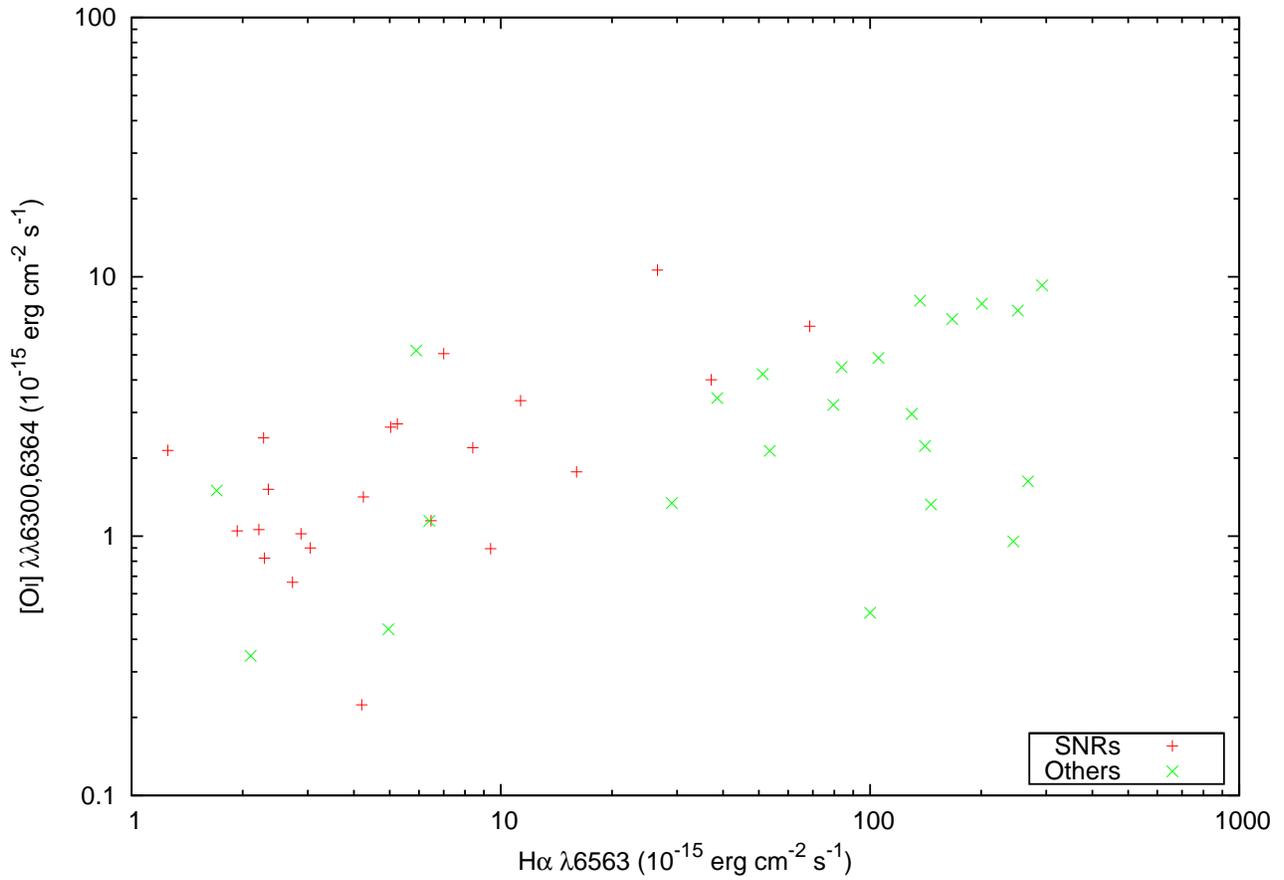}}
  \caption{A plot of [\ion{O}{i}] vs. \Ha\ flux values for the objects we label as SNRs and ``other objects.'' The scales were made logarithmic to make the data points more visible. The plot shows a consistent value of [\ion{O}{i}] emission from both object groups, with greater \Ha\ emission form the ``other'' objects. Because [\ion{O}{i}] flux is associated with SNR shock fronts it may be possible that the emission from the ``other objects'' is caused by shock fronts created by SNRs that are visible only in radio, being buried within \HII\ regions.}
  \label{fig:hao1}
\end{figure*}

{Figure~\ref{fig:has2} shows a plot of the \linrat\ values for all objects which returned data. We can see a trend for objects which are below 0.4 to have greater \Ha\ emissions, lending support to these objects appearing as \HII\ regions rather than SNRs at optical wavelengths. Figure~\ref{fig:han2} shows a plot of [\ion{N}{ii}] flux against \Ha\ flux. This plot shows a fairly consistent ratio of [\ion{N}{ii}]$_{Total}$:\Ha\ of 0.3 across both groups of objects. Figure~\ref{fig:hao1} shows a plot of [\ion{O}{i}] flux against \Ha\ flux. The plot shows a consistent value of [\ion{O}{i}] emission from both object groups, with greater \Ha\ emission from the ``other'' objects. Because [\ion{O}{i}] flux is associated with SNR shock fronts it may be possible that the emission from the ``other sources'' is caused by shock fronts created by SNRs that are visible only in radio, being buried within \HII\ regions.}

\subsection{Notes on Individual Radio Objects.}

\subsubsection{J005431-373825.}

The radio source J005431-373825 is associated with the optical object N300-S6 and thus has both radio and optical emission. The line ratio of 0.69 confirms this object as a SNR. This SNR also shows hard X-ray emission (C05).

\subsubsection{J005438-374144.}

There is a faint optical object at the position of this radio source which was previously classified as SNR by P04. The spectrum shows a \linrat\ ratio of 0.17 with an error of 0.07. On the basis of this we do not confirm this object as a SNR although this source may be a radio SNR hidden within an \HII\ region.

\subsubsection{J005440-374049.}

The radio source J005440-374049 (\NS10) was observed three times (once more after the main runs, with $PA=\Deg{45}$) because of the interesting structure seen on the DSS2-Red. All three observations return a line ratio of $<0.4$. Our diameter measurement for this object reveal it to be 63~pc, significantly larger than BL97 who describe it as very compact (16~pc). Based on the proposed radio and optical identification from P04 and BL97, our X-ray identification (also see Sect~\ref{xray}) and the \linrat\ of $<0.4$, we classify this object as a candidate SNR. 

\subsubsection{J005441-373348.}

We did not see any optical emission from this source but it was observed in X-rays as XMM4 by P04 and as source \#54 in \citet{Carpano2005}. As these papers have noted, this object is probably a background source.

\subsubsection{J005442-374313 and J005443-374311.}

The sources \ATCA{5442}{4313}, \ATCA{5443}{4311} and \NS11, are in the same line of sight and may be located near to each other. They also have similar optical spectra and \linrat\ ratios (Table~\ref{tab:6}). \NS11 was observed in both P00 and BL97 and identified as an SNR in BL97. P00 searched for X-ray and radio emissions from this source but no definitive counterparts could be identified. X-ray source \#161 in the catalog given by \citet{Carpano2005} is nearby, but not within the positional uncertainties given in the respective papers. With these data, these objects remain candidate ``radio'' or ``optical'' SNRs and their physical association is still questionable. Table~\ref{tab:6} gives the positional agreement between these sources and source \#161 from \citet{Carpano2005} X-ray Source Catalog.

\subsubsection{J005450-374030 and J005450-374022.}

Radio sources J005450-374030 and J005450-374022 are 10.3 arcseconds ($\sim$100~pc) from each other. The \linrat\ for these objects are 0.32$\pm$0.12 and \linebreak{}0.38$\pm$0.31 respectively. They may be two radio emission regions within the same (perhaps older and larger) SNR or simply two neighboring candidate radio SNRs.

\subsubsection{ J005501-373829.}

J005501-373829 is on the western edge of a large region of optical emission (dim in DSS2-Red but bright in DSS; perhaps indicating a strong component of reflected emission). This object has radio (P04) emission but its optical emission has \linrat\ of 0.35$\pm$0.12. We also note the proximity of the OB association AS\_082 \citep{Pietrzynski2001} which is cross-referenced by those authors to \HII\ region 115 in the \citet{Deharveng1988} catalog. 

\subsubsection{J005515-374439.}

The radio source J005515-374439 is associated with \NS26 and has \linrat\ of 0.86$\pm$0.67. We confirm this source as a bona-fide SNR. It is on the edge of a small, circular, faint object visible in both DSS2-Red and DSS images. This SNR has proximity to OB association AS\_107, \HII\ region 141 and also to P04's XMM9 X-ray source. Also, \citet{Carpano2005} lists \NS26 as a soft X-ray source (\#34). From our complementary HRC X-ray observations of this object we confirmed a discrete X-ray source within 2~arc seconds of the radio position. For more details see Sect~\ref{xray}. 

\subsubsection{ J005533-374314.}

J005533-374314 (associated with \NS28) has a measured \linrat\ of 0.45$\pm$0.15 and an optical counterpart with an
estimated size of 63~pc, which in both DSS2-Red and DSS has a two-lobed shape. The radio emission is southeast of the center of the lobes. \ATCA{5533}{4314} may be linked to the OB association AS\_113 and \HII\ region 159. C05 found an X-ray source at this location (\#151). Better optical imaging and further X-ray observations of this SNR will prove beneficial.

\subsection{Sources with No Measurable Spectrum}

We did not detect a measurable level of \Ha\ and/or [\ion{S}{ii}] flux against the background noise from the following sources: \ATCA{5423}{3648}, \ATCA{5521}{4609}, \ATCA{5523}{4632}, \ATCA{5525}{3653}, \ATCA{5528}{4903}, and \ATCA{5541}{4033}. The nature of these sources is thus uncertain.

P04 suggests the two radio sources \ATCA{5423}{3648} and \ATCA{5528}{4903} as low confidence SNR candidates. The other P04 sources in the above list have radio emission which would not exclude them as possible SNRs. Given the data and discussion in P04, failure to detect an optical emission spectrum typical of SNRs is consistent with the P04 expectations.

\subsection{The Multi-Wavelength Properties of the SNRs}

\begin{figure*}[!h]
%  \centerline{\includegraphics{VennDia.eps}}
  \centerline{\includegraphics[angle=0,width=1.0\textwidth]{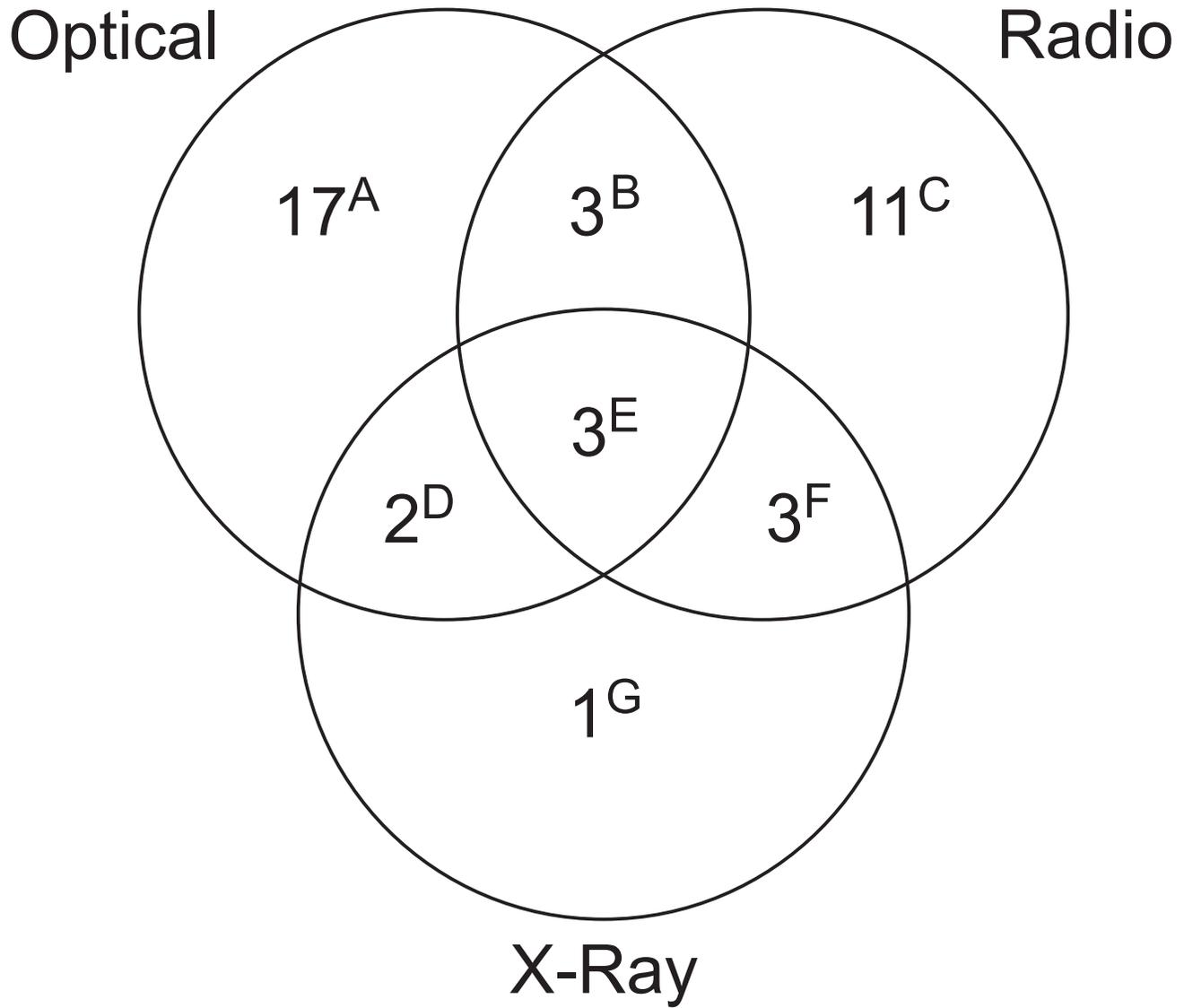}}
%  \epsscale{.60}
%  \plotone{Fig4.eps}
  \caption{Venn diagram showing the intersection of selected sets of SNRs for NGC\,300. The letter superscript on each value corresponds to the ``Venn Region'' column of Table~\ref{tab:7}.}
%  \label{fig:Venn}
  \label{fig:5}
\end{figure*}

In P00, P04, \citet{2010ApJS..187..495L}, \citet{2007AJ....133.1361P} and Pannuti et al (2010, submitted), Venn diagrams were used to show the number of ``radio,'' ``optical,'' and ``X-ray'' SNRs in NGC\,300, M\,33 and NGC\,7793. In Figure~\ref{fig:5}, we create a new diagram using results from this paper and papers P04, P00 and \citet{Carpano2005, Carpano2006a}. The SNRs included within each region are listed in Table~\ref{tab:7}.

The majority of the SNRs (29 out of 40) are identified with emission in only one wavelength region.  We note also the low numbers of multiple-frequency and high-energy SNRs; there are only six SNRs detected in both radio and optical wavelengths and only three detected in all three wavelength regions. One SNR was detected in X-ray wavelengths only, and nine SNRs detected in X-ray wavelengths.

Our results are somewhat similar to the M~33 findings of \citet{2010ApJS..187..495L} and in NGC\,7793 (Pannnuti et al 2010, submitted) where a large number of optical-only SNRs are identified. This may imply a possible observational bias towards the optical techniques of SNR detection in external galaxies where the resolution (and sensitivity) may play a dominant role in SNR identification. Further to this argument, \citet{2010ApJ...710..964D, 2010ApSS.tmp..148D} showed that the \linrat\ is a very ``sharp'' tool in the case of two other nearby galaxies (M\,83 and NGC\,4214) where they use very advanced Hubble Space Telescope (HST) imaging and spectroscopy. Therefore, we note that the population in the Venn cells is probably of little astrophysical consequence as it is determined by selection effects and the sensitivity limits of the present radio, optical and X-ray surveys. Details of these limits are given in P04, \citet{2007AJ....133.1361P} and BL97, and by C05 for the X-ray observations reported here.

It is accepted \citep{Filipovic1998} that all SNRs will emit some thermal radiation at all of these wavelengths. In addition, most SNRs exhibit non-thermal properties in radio and a handful of SNRs in X-rays. However, the amount of emission at each band is the subject of numerous studies, and it is generally understood that the emission is determined by the SN event itself and the ISM in which it occurs. These two variables lead to a wide range of observed structural and emission properties when SNRs are studied over the full electromagnetic spectrum, and no standard spectrum over all wavelengths exists for SNRs.

Compounding these variations are the selection effects that have been introduced. The radio observations of P04 have preferentially selected all sources with spectral index below $-0.2$ (a safer spectral index selection cut-off value would be $-0.5$), which probably excludes half of the SNRs available at these wavelengths, and selects preferentially the non-thermal objects. Similarly, the filter-based optical selection of BL97 preferentially selects objects with strong [\ion{S}{ii}]. This again is a safe choice for locating SNRs, but will miss older objects or the [\ion{O}{iii}] dominated objects that are perhaps present at the 10 to 15\% level \citep{Stupar2008}. It is clear that further studies conducted with instruments featuring improved sensitivity will result in the detection of more SNRs within each wavelength region thus increasing the populations of individual cells of the Venn diagram.

To complete this study we investigated the dependence of the \linrat\ with distance from the galactic nucleus. We found no statistically-significant dependence.

%==================================================================================================

\section{Summary}

It is not surprising that we found only a few radio SNR candidates exhibiting the established \linrat. At the distance to this galaxy (2.1~Mpc), with this slit width (1\arcsec) and the seeing spread along the decker, we are sampling about 100 square parsecs per ``line.'' Thus, each spectral line contains the light from $\sim$50\,000 field stars as well as the emission from the nebula. Although \linrat~$\geq0.4$ may be fine for galaxies of the Local Group, it may not be the best tool for detecting distant extra-galactic SNRs unless we use the highest resolution telescopes such as the HST. We summarize our findings as follows:

\begin{enumerate}
\item We obtained moderate resolution optical spectra of 51 nebular objects in the Sculptor Group galaxy NGC~300. Of the 51 objects, 4 were proposed as SNR candidates in both optical and radio observations, 24 are proposed SNR candidates from optical spectra and 23 are SNR candidates from radio spectra.
\item We find 22 objects meeting the accepted \linebreak{}\linrat\ as SNRs, with the nature of the remaining 29 objects either unclear or unknown. A slight bias (of order 10\%) appears between our \linrat\ for the optical candidates compared to a previous observation of these same objects (BL97).
\item We created a Gaussian fit of the image of the nebular object across the spectroscope's slit to estimate the diameter of the candidate SNRs. Comparison of our diameter measurements does not show good correlation with previous size estimates (BL97).
\item We also found 31 X-ray sources in this galaxy, 2 of which are positionally linked to nebulae meeting the accepted \linrat\ ratio as SNRs.
\end{enumerate}

%==================================================================================================

\section*{Acknowledgments}
We thank the Australian National University Research School of Astronomy and Astrophysics (RSAA) for the use of their 2.3~m Advanced Technology Telescope. We used the NOAO (National Optical Astronomical Observatory) IRAF (Image Reduction and Analysis Facility) software package, with Brent Miszalski's extractor task for IRAF, and the Figaro software package from the Starlink Project. IRAF is distributed by the National Optical Astronomy Observatories, which are operated by the Association of Universities for Research in Astronomy, Inc. (AURA), under cooperative agreement with the National Science Foundation. We also used the KARMA software package.

%\nocite{*}
\bibliographystyle{spr-mp-nameyear-cnd}
\bibliography{PaperV10.2}

%\end{document}

%%%%%%%%%%%%%%%%%%%%%%%%%%%%%%%%%%%%%%%%%%%%%%%%%%%%%%%%%%%%%%%%%%%%%%%%
\newpage

% Fig 1

\clearpage

% Fig 2

\clearpage

% Fig 3

\clearpage

% Fig 4

\clearpage

% Fig
%

% Fig
%

% Fig

\clearpage

% Fig 5

\clearpage

%%%%%%%%%%%%%%%%%%%%%%%%%%%%%%%%%%%%%%%%%%%%%%%%%%%%%%%%%%%%%%%%%%%%%%

%Table 1
\begin{deluxetable}{lll}
  \tabletypesize{\small}
  \tablecaption{Gross Properties of NGC\,300\label{tab:1}}
  \tablewidth{0pt}
  \tablehead{\colhead{Property} & \colhead{Value} & \colhead{Reference}}
  \startdata
  Hubble Type                & SA(s)d                                & \citet{Tully1988} \\
                             &                                       & \citet{deVaucouleurs1991} \\
  R.A. (J2000.0)             & \RA{00}{54}{53.48}                    & NED \\
  Dec. (J2000.0)             & \DEC{-37}{41}{03.8}                   & NED \\
  Galactic Latitude          & \Deg{-77.17}                          & NED \\
  Radial Velocity            & 144 km/s (Solar)                      & \citet{Puche1990} \\
                             &                                       & \citet{Karach2003} \\
  Inclination                & \Deg{46}                              & \citet{Tully1988} \\
                             & \Deg{42.6}                            & \citet{Puche1990} \\
  Distance                   & 2.1 Mpc                               & \citet{Freedman1992} \\
                             & 2.02 Mpc                              & \citet{Freedman2001} \\
                             & 1.88 Mpc                              & \citet{Bresolin2005}; \\
                             &                                       & \citet{Gieren2005} \\
  Observed Diameter ($D_{25}$) & 20.2 arcmin                         & \citet{Tully1988} \\
  Observed Diameter (UV isophotes) & 21.9 x 15.5 arcmin              & \citet{dePaz2007} \\
  Galaxy Diameter            & 22.6 kpc, at 2.1 Mpc                  & Based on \citet{dePaz2007} \\
  Mass (\HI)                 & $2.4\times10^9\;$\msun                & \citet{Tully1988} \\
  $N_H$ Column Density       & $2.97\times10^{20}\;\mathrm{cm}^{-2}$ & \citet{Read1997} \\
  \enddata
  \tablecomments{Note. NED = NASA/IPAC Extragalactic Database (http://nedwww.ipac.caltech.edu/).}
\end{deluxetable}

\clearpage

% Table 2
\begin{deluxetable}{lcc|lcc}
  \tabletypesize{\small}
  \tablecolumns{6}
  \tablecaption{Positions (J2000.0) of the selected 51 SNRs and SNR candidates in the NGC\,300\label{tab:2}}
  \tablewidth{0pt}
  \tablehead{
  \colhead{1} & \colhead{2} & \colhead{3} & \colhead{1} & \colhead{2} & \colhead{3} \\
  \colhead{Object Name} & \multicolumn{2}{c}{Position (J2000.0)}  & \colhead{Object Name} & \multicolumn{2}{c}{Position (J2000.0)} \\
   & \colhead{RA (h m s)} & \colhead{Dec (\Deg{}\ \arcmin\ \arcsec)} &  & \colhead{RA (h m s)} & \colhead{Dec (\Deg{}\ \arcmin\ \arcsec)}}
  \startdata
  \ATCA{5423}{3648}  & 00 54 23.84 & $-37$ 36 48.4             & \NS1    & 00 54 19.21 & $-37$ 37 23.96 \\
  \ATCA{5431}{3825}  & 00 54 31.91 & $-37$ 38 25.9             & \NS2    & 00 54 21.85 & $-37$ 40 27.11 \\
  \ATCA{5438}{4144}  & 00 54 38.16 & $-37$ 41 44.2             & \NS3    & 00 54 28.86 & $-37$ 41 53.32 \\
  \ATCA{5438}{4240}  & 00 54 38.49 & $-37$ 42 40.5             & \NS4    & 00 54 30.62 & $-37$ 40 53.75 \\
  \ATCA{5439}{3543}  & 00 54 39.61 & $-37$ 35 43.4             & \NS5    & 00 54 30.99 & $-37$ 37 33.96 \\
  \ATCA{5440}{4049}  & 00 54 40.68 & $-37$ 40 49.7             & \NS6    & 00 54 31.91 & $-37$ 38 25.68 \\
  \ATCA{5441}{3348}  & 00 54 41.05 & $-37$ 33 48.9             & \NS7    & 00 54 33.17 & $-37$ 40 16.90 \\
  \ATCA{5442}{4313}  & 00 54 42.70 & $-37$ 43 13.3             & \NS8    & 00 54 38.17 & $-37$ 41 14.88 \\
  \ATCA{5443}{4311}  & 00 54 43.11 & $-37$ 43 11.0             & \NS9    & 00 54 40.20 & $-37$ 41 02.12 \\
  \ATCA{5445}{3847}  & 00 54 45.39 & $-37$ 38 47.1             & \NS10   & 00 54 40.87 & $-37$ 40 48.73 \\
  \ATCA{5450}{4030}  & 00 54 50.28 & $-37$ 40 30.0             & \NS11   & 00 54 42.54 & $-37$ 43 14.16 \\
  \ATCA{5450}{3822}  & 00 54 50.30 & $-37$ 38 22.4             & \NS12   & 00 54 43.86 & $-37$ 43 39.08 \\
  \ATCA{5450}{4022}  & 00 54 50.73 & $-37$ 40 22.2             & \NS13   & 00 54 46.60 & $-37$ 39 44.32 \\
  \ATCA{5451}{3826}  & 00 54 51.16 & $-37$ 38 26.1             & \NS14   & 00 54 47.15 & $-37$ 41 07.63 \\
  \ATCA{5451}{3939}  & 00 54 51.79 & $-37$ 39 39.6             & \NS15   & 00 54 53.32 & $-37$ 38 48.24 \\
  \ATCA{5500}{4037}  & 00 55 00.58 & $-37$ 40 37.4             & \NS16   & 00 54 54.46 & $-37$ 40 35.46 \\
  \ATCA{5501}{3829}  & 00 55 01.49 & $-37$ 38 29.9             & \NS17   & 00 54 56.68 & $-37$ 43 57.70 \\
  \ATCA{5503}{4246}  & 00 55 03.50 & $-37$ 42 46.0             & \NS18   & 00 55 01.39 & $-37$ 39 18.17 \\
  \ATCA{5503}{4320}  & 00 55 03.66 & $-37$ 43 20.1             & \NS19   & 00 55 05.41 & $-37$ 41 21.04 \\
  \ATCA{5512}{4140}  & 00 55 12.70 & $-37$ 41 40.3             & \NS20   & 00 55 05.68 & $-37$ 46 13.35 \\
  \ATCA{5515}{4439}  & 00 55 15.40 & $-37$ 44 39.2             & \NS21   & 00 55 07.15 & $-37$ 39 15.17 \\
  \ATCA{5521}{4609}  & 00 55 21.35 & $-37$ 46 09.6             & \NS22   & 00 55 07.50 & $-37$ 40 43.20 \\
  \ATCA{5523}{4632}  & 00 55 23.95 & $-37$ 46 32.4             & \NS23   & 00 55 09.10 & $-37$ 39 32.61 \\
  \ATCA{5525}{3653}  & 00 55 25.82 & $-37$ 36 53.8             & \NS24   & 00 55 09.48 & $-37$ 40 46.21 \\
  \ATCA{5528}{4903}  & 00 55 28.25 & $-37$ 49 03.3             & \NS25   & 00 55 10.68 & $-37$ 41 27.13 \\
  \ATCA{5533}{4314}  & 00 55 33.87 & $-37$ 43 14.6             & \NS26   & 00 55 15.46 & $-37$ 44 39.11 \\
  \ATCA{5541}{4033}  & 00 55 41.94 & $-37$ 40 33.5             & \NS27   & 00 55 17.54 & $-37$ 44 36.65 \\
                     &             &                           & \NS28   & 00 55 33.76 & $-37$ 43 13.13 \\
  \enddata
\end{deluxetable}

\clearpage

% Table 3
\begin{deluxetable}{ccccccccccc}
  \tabletypesize{\tiny}
  \rotate
  \tablecaption{Previous Observations of the Selected Objects\label{tab:3}}
  \tablewidth{0pt}
  \tablecolumns{11}
  \tablehead{
  \colhead{1} & \colhead{2} & \colhead{3} & \colhead{4} & \colhead{5} & 
  \colhead{6} & \colhead{7} & \colhead{8} & \colhead{9} & \colhead{10} & \colhead{11} \\
  
  \multicolumn{2}{c}{Object Designation} & 
  \multicolumn{3}{c}{Radio Observations} & 
  \multicolumn{4}{c}{X-ray Observations} & 
  \multicolumn{2}{c}{Optical Observation (BL97)} \\
  
  \colhead{Optical} & \colhead{Radio} & \colhead{Spectral} & \colhead{Spectral} & \colhead{Proposed} & 
  \colhead{RPS97 Label} & \colhead{RP01 Label} & \colhead{P04 Label} & \colhead{C05 Label} & 
  \colhead{DSS2-Red} & \colhead{[\ion{S}{ii}]:\Ha} \\
  
  &  & \colhead{Index (P00)} & \colhead{Index (P04)} & \colhead{Object (P04)} & \colhead{Data} & 
  \colhead{Data} & \colhead{Data} & \colhead{Data}
  }

  \startdata
  \multicolumn{11}{c}{\tiny SNRs} \\
  \hline
  \NS1 & & & & & & & & & No & 0.68 \\
  \NS2 & & & & & & & & \#79 & Yes & 0.77 \\
  \NS4 & & & & & & & & & Yes & 0.99 \\
  \NS5 & & & & & & & & & Yes & 0.68 \\
  \NS6 & \ATCA{5431}{3825} & & & SNR & \#1 & P29 & XMM2 & & Yes & 0.96 \\
  \NS7 & & & & & & & & & Yes & 0.72 \\
  \NS8 & & & & & & & & & Yes & 0.61 \\
  \NS9 & & & & & & & & & Yes & 0.67 \\
  \NS12 & & & & & & & & & Yes & (0.52) \\
  \NS13 & & & & & & & & & Yes & 0.98 \\
  \NS14 & & & & & & & & & Yes & 0.83 \\
  \NS15 & & & & & & & & & Yes & 0.74 \\
  \NS16 & & & & & & & & & Yes & (0.70) \\
  \NS17 & & & & & & & & & Yes & (0.69) \\
  \NS19 & & & & & & & & \#123 & Yes & 0.90 \\
  \NS20 & & & & & & & & & Yes & 1.00 \\
  \NS22 & & & & & & & & & Yes & 0.57 \\
  \NS24 & & & & & & & & & No & 0.83 \\
  \NS26 & \ATCA{5515}{4439} & & & SNR/HII & \#10 & P49 & XMM9 & \#34 & Yes & 1.03 \\
  \NS25 & & & & & & & & & Yes & 0.80 \\
  \NS27 & & & & & & & & & Yes & 1.05 \\ 
  \NS28 & \ATCA{5533}{4314} & & & SNR/HII & & & & \#151 & Yes & 0.72 \\

  \hline
  \multicolumn{11}{c}{\tiny Other Objects (\HII\ Regions?)} \\
  \hline

  & \ATCA{5438}{4144} & $<1.1$ & $-0.8$ & SNR/HII & & & & & Yes \\
  & \ATCA{5438}{4240} & $<1.5$ & & snr/HII & & & & & Yes \\
  & \ATCA{5439}{3543} & & $-0.4$ & snr$\dagger$ & & & & & No \\
  & & & & & 36.66, 82.8, 0.10 & 0.87, $-0.33$ & 1.27, $-0.63$, $-0.99$ & $-0.17$, $-0.58$ & & \\
  & & & & & & 4.8, 25.4 & 1.63, $-0.58$, $-1.00$ & 1.03, 5.03 & & \\
  & \ATCA{5441}{3348} & & & bkg/snr & & & XMM4 & & Yes \\
  & & & & & & & 1.39, $+0.84$, $+0.94$ & & Yes \\
  & & & & & & & 0.72, $+0.31$, $+0.62$ & & Yes \\
  & \ATCA{5442}{4313} & & $-0.9$ & SNR/HII & & & & & Yes & (0.66) \\
  & \ATCA{5443}{4311} & $-0.6$ & $-0.6$ & SNR/HII & & & & & Yes \\
  & \ATCA{5445}{3847} & & $-0.3$ & SNR/HII & & & & & Yes \\
  & \ATCA{5450}{4030} & $-0.6$ & $-0.5$ & SNR/HII & & & & & Yes \\
  & \ATCA{5450}{3822} & & $-0.2$ & SNR/HII & & & & & Yes \\
  & \ATCA{5450}{4022} & & $-0.3$ & SNR/HII & & & & & Yes \\
  & \ATCA{5451}{3826} & & $-1.2$ & SNR/HII & & & & & Yes \\
  & \ATCA{5451}{3939} & $-0.4$ & $-0.1$ & SNR/HII & & & & & Yes \\
  & \ATCA{5500}{4037} & & $-0.4$ & SNR/HII & & & & & Yes \\
  & \ATCA{5501}{3829} & & $-0.9$ & SNR & & & & & Yes \\
  & \ATCA{5503}{4246} & $-0.2$ & $-0.4$ & SNR/HII & & & & & Yes \\
  & \ATCA{5503}{4320} & $-1.0$ & $-0.7$ & SNR/HII & & & & & Yes \\
  & \ATCA{5512}{4140} & $-0.4$ & $-0.7$ & SNR/HII & & & & & Yes \\
  \NS3 & & & & & & & & & No & (0.40) \\
  \NS10 & \ATCA{5440}{4049} & & $-0.5$ & SNR & \#4 & P38 & XMM3 & \#12 & Yes & 0.47 \\
  \NS$11^a$ & \ATCA{5442}{4313?} & & $-0.9$ & SNR/HII & & & & & Yes & (0.66) \\
  \NS18 & & & & & & & & & Yes & 0.71 \\
  \NS21 & & & & & & & & & Yes & (0.59) \\
  \NS23 & & & & & & & & & Yes & (0.46) \\

  \hline
  \multicolumn{11}{c}{\tiny No Signal} \\
  \hline

  & \ATCA{5423}{3648} & & $-0.7$ & snr$\dagger$ & & & & & No \\
  & \ATCA{5521}{4609} & & $-1.0$ & bkg/snr & & & & & No \\
  & \ATCA{5523}{4632} & & $-0.9$ & bkg/snr & & & & & No \\
  & \ATCA{5525}{3653} & & $-1.0$ & bkg/snr & & & & & No \\
  & \ATCA{5528}{4903} & & $-0.6$ & snr$\dagger$ & & & & & No \\
  & \ATCA{5541}{4033} & & & snr & & & XMM10 & & No \\
  & & & & & & & 2.35, $-0.76$, $-0.88$ & & No \\
  & & & & & & & 1.08, $-0.88$, $-0.93$ & & No \\

  \enddata
\end{deluxetable}

\clearpage

% Table 4
\begin{deluxetable}{cccccccc}
  \tabletypesize{\tiny}
  \rotate
  \tablecaption{The Integrated Line Flux Density Measurements.\label{tab:4}}
  \tablewidth{0pt}
  \tablehead{
  \colhead{1} & \colhead{2} & \colhead{3} & \colhead{4} & \colhead{5} & \colhead{6} & \colhead{7} & \colhead{8} \\

  \multicolumn{2}{c}{\tiny Designation} & \colhead{\Ha\ \wl{6563}} & \colhead{[\ion{N}{ii}] total} & 
  \colhead{[\ion{S}{ii}] total} &  & \colhead{[\ion{O}{i}] total} & \colhead{Diameter} \\

  \colhead{Radio} & \colhead{Optical} & \colhead{($10^{-15}$ \Units{ergs\;cm^{-2}\;s^{-1}})} & 
  \colhead{($10^{-15}$ \Units{ergs\;cm^{-2}\;s^{-1}})} & \colhead{($10^{-15}$ \Units{ergs\;cm^{-2}\;s^{-1}})} & 
  \colhead{\linrat} & \colhead{($10^{-15}$ \Units{ergs\;cm^{-2}\;s^{-1}})} & \colhead{(pc)}
  }
  \startdata
  \multicolumn{7}{c}{\tiny SNRs} \\
  \hline
  & \NS1\tablenotemark{a} & 1.3 & 0.25 & 0.57 & 0.46$\pm$0.29 & 2.1 & 38 \\
  %Map 29
  & \NS2 & 11 & 3.7 & 8.1 & 0.72$\pm$0.39 & 3.3 & 69 \\
  %Map 31
  & \NS4 & 7.0 & 3.0 & 6.5 & 0.93$\pm$0.04 & 5.1 & 150 \\
  %Map 32
  & \NS5 & 8.4 & 2.4 & 4.7 & 0.56$\pm$0.47 & 2.2 & \nosig\tablenotemark{b} \\
  %Map 2
  \ATCA{5431}{3825} & \NS6 & 4.2 & 1.2 & 2.9 & 0.69 & 0.22 & 44 \\
  %Map 33
  & \NS7 & 5.2 & 2.1 & 3.0 & 0.57$\pm$0.41 & 2.7 &  31 \\
  %Map 34
  & \NS8 & 5.0 & 1.9 & 2.9 & 0.58$\pm$0.06 & 2.6 & 49 \\
  %Map 35
  & \NS9 & 37 & 12 & 20 & 0.53$\pm$0.23 & 4.0 & 83 \\
  %Map 37
  & \NS12 & 2.3 & 1.2 & 1.7 & 0.73$\pm$0.27 & 2.4 & 22 \\
  %Map 38
  & \NS13 & 1.9 & 1.1 & 1.8 & 0.91$\pm$0.05 & 1.0 & 35 \\
  %Map 39
  & \NS14 & 2.3 & 1.8 & 2.5 & 1.08$\pm$0.24 & 1.5 & 41 \\
  %Map 40
  & \NS15 & 3.0 & 1.0 & 1.7 & 0.57$\pm$0.39 & 0.90 & 12 \\
  %Map 41
  & \NS16 & 2.2 & 2.0 & 2.1 & 0.94$\pm$0.06 & 1.1 & 57 \\
  %Map 42
  & \NS17 & 4.2 & 1.6 & 4.1 & 0.96$\pm$0.15 & 1.4 & 65 \\
  %Map 44
  & \NS19 & 9.4 & 3.8 & 6.6 & 0.70$\pm$0.42 & 0.90 & 30 \\
  %Map 45
  & \NS20 & 2.3 & 0.71 & 1.8 & 0.79$\pm$0.11 & 0.28 & 48 \\
  %Map 47
  & \NS22 & 2.9 & 1.1 & 1.3 & 0.46$\pm$0.38 & 1.0 & 75 \\
  %Map 49
  & \NS24 & 2.7 & 0.86 & 1.7 & 0.64$\pm$0.13 & 0.66 & 100 \\
  %Map 50
  & \NS25 & 16 & 4.7 & 8.7 & 0.54$\pm$0.40 & 1.8 & 80 \\
  %Map 21
  \ATCA{5515}{4439} & \NS26\tablenotemark{a} & 26 & 11 & 23 & 0.86$\pm$0.67 & 11 & 31 \\
  %Map 51
  & \NS27 & 6.5 & 1.7 & 4.2 & 0.64$\pm$0.48 & 1.1 & 66 \\
  %Map 26
  \ATCA{5533}{4314} & \NS28\tablenotemark{a} & 69 & 13 & 31 & 0.45$\pm$0.15 & 6.5 & 63 \\
  
  \hline
  \multicolumn{7}{c}{\tiny Other Objects (\HII\ Regions?)} \\
  \hline

  %Map 3
  \ATCA{5438}{4144} & & 80 & 15 & 13 & 0.17$\pm$0.07 & 3.2 \\
  %Map 4
  \ATCA{5438}{4240} & & 39 & 8.5 & 10 & 0.25$\pm$0.02 & 3.4 \\
  %Map 5
  \ATCA{5439}{3543} & & 5.9 & 1.4 & 1.6 & 0.27 & 5.2 \\
  %Map 6
  \ATCA{5440}{4049} & \NS10 & 105 & 32 & 36 & 0.35$\pm$0.15 & 4.9 \\
  %Map 7
  \ATCA{5441}{3348}\tablenotemark{a} & & 1.7 & 0.34 & 0.62 & 0.36 & 1.5 \\
  %Map 8
  \ATCA{5442}{4313} & & 250 & 48 & 49 & 0.19$\pm$0.07 & \\
  %Map 9
  \ATCA{5443}{4311} & & 290 & 58 & 53 & 0.18$\pm$0.09 & \\
  %Map 10
  \ATCA{5445}{3847} & & 130 & 20 & 14 & 0.11$\pm$0.03 & \\
  %Map 11
  \ATCA{5450}{4030} & & 140 & 46 & 43 & 0.32$\pm$0.12 & 8.1 & 24 \\
  %Map 12
  \ATCA{5450}{3822} & & 200 & 40 & 44 & 0.22$\pm$0.14 & 7.9 \\
  %Map 13
  \ATCA{5450}{4022} & & 84 & 28 & 32 & 0.38$\pm$0.31 & 4.5 & 130 \\
  %Map 14
  \ATCA{5451}{3826} & & 140 & 30 & 36 & 0.26$\pm$0.16 & 2.2 \\
  %Map 15
  \ATCA{5451}{3939}\tablenotemark{a} & & 100 & 16 & 10 & 0.10$\pm$0.18 & 0.51 \\
  %Map 16
  \ATCA{5500}{4037} & & 53 & 17 & 13 & 0.25$\pm$0.09 & 2.1 \\
  %Map 17
  \ATCA{5501}{3829} & & 6.4 & 1.5 & 2.3 & 0.35$\pm$0.12 & 1.1 & 31 \\
  %Map 18
  \ATCA{5503}{4246} & & 150 & 24 & 20 & 0.14$\pm$0.04 & 1.3 \\
  %Map 19
  \ATCA{5503}{4320} & & 270 & 53 & 41 & 0.15$\pm$0.08 & 1.6 \\
  %Map 20
  \ATCA{5512}{4140} & & 240 & 27 & 19 & 0.08$\pm$0.02 & 0.95 \\
  %Map 30
  & \NS3 & 51 & 10 & 12 & 0.24$\pm$0.31 & 4.2 & 26 \\
  %Map 6
  \ATCA{5440}{4049} & \NS10 & 105 & 32 & 36 & 0.35$\pm$0.15 & 4.9 \\
  %Map 36
  & \NS11 & 1670 & 39 & 50 & 0.30$\pm$0.12 & 6.9 & 150 \\
  %Map 43
  & \NS18\tablenotemark{a} & 5.0 & 1.5 & 1.6 & 0.32$\pm$0.32 & 0.44 & 69 \\
  %Map 46
  & \NS21 & 2.1 & 0.72 & 0.78 & 0.37$\pm$0.30 & 0.35 & 41 \\
  %Map 48
  & \NS23 & 29 & 7.2 & 9.1 & 0.31$\pm$0.08 & 1.3 & 43 \\

  \hline
  \multicolumn{7}{c}{\tiny No Signal\tablenotemark{c}} \\
  \hline

  %Map 1
  \ATCA{5423}{3648} \\
  %Map 22
  \ATCA{5521}{4609} \\
  %Map 23
  \ATCA{5523}{4632} \\
  %Map 24
  \ATCA{5525}{3653} \\
  %Map 25
  \ATCA{5528}{4903} \\
  %Map 27
  \ATCA{5541}{4033} \\

  \enddata
  \tablenotetext{a}{Very low faint spectrum recorded.}
  \tablenotetext{b}{Gaussian fit returned value too small to be deconvolved with reference star.}
  \tablenotetext{c}{No observed spectral lines.}
\end{deluxetable}

\clearpage

% Table 5
\begin{deluxetable}{cccccc}
  \tabletypesize{\tiny}
  \tablecaption{Summary of results\label{tab:5}}
  \tablewidth{0pt}
  \tablehead{
  \colhead{1} & \colhead{2} & \colhead{3} & \colhead{4} & \colhead{5} & \colhead{6} \\
  \colhead{Optical} & \colhead{Radio} & \colhead{Object} & \colhead{Object} & & \colhead{Diameter} \\
  \colhead{object}  & \colhead{object} & \colhead{type (P04)} & \colhead{type (BL97)} & \colhead{\linrat} & \colhead{(pc)}
  }
  \startdata
  \multicolumn{6}{c}{\tiny SNRs} \\
  \hline
  \NS1  &                   &              & SNR & 0.46$\pm$0.29   & 38 \\ %Map 28
  \NS2  &                   &              & SNR & 0.72$\pm$0.39   & 69 \\ %Map 29
  \NS4  &                   &              & SNR & 0.93$\pm$0.04   & 150 \\ %Map 31
  \NS5  &                   &              & SNR & 0.56$\pm$0.47   & \nosig\tablenotemark{a} \\ %Map 32
  \NS6  & \ATCA{5431}{3825} & SNR          & SNR & 0.69            & 44 \\ %Map 2
  \NS7  &                   &              & SNR & 0.57$\pm$0.41   & 31 \\ %Map 33
  \NS8  &                   &              & SNR & 0.58$\pm$0.06   & 49 \\ %Map 34
  \NS9  &                   &              & SNR & 0.53$\pm$0.23   & 83 \\ %Map 35
  \NS12 &                   &              & SNR & 0.73$\pm$0.27   & 22 \\ %Map 37
  \NS13 &                   &              & SNR & 0.91$\pm$0.05   & 35 \\ %Map 38
  \NS14 &                   &              & SNR & 1.08$\pm$0.24   & 41 \\ %Map 39
  \NS15 &                   &              & SNR & 0.57$\pm$0.39   & 12 \\ %Map 40
  \NS16 &                   &              & SNR & 0.94$\pm$0.06   & 57 \\ %Map 41
  \NS17 &                   &              & SNR & 0.96$\pm$0.15   & 65 \\ %Map 42
  \NS19 &                   &              & SNR & 0.70$\pm$0.42   & 30 \\ %Map 44
  \NS20 &                   &              & SNR & 0.79$\pm$0.11   & 48 \\ %Map 45
  \NS22 &                   &              & SNR & 0.46$\pm$0.38   & 75 \\ %Map 47
  \NS24 &                   &              & SNR & 0.64$\pm$0.13   & 100 \\ %Map 49
  \NS25 &                   &              & SNR & 0.54$\pm$0.40   & 80 \\ %Map 50
  \NS26 & \ATCA{5515}{4439} & SNR/HII      & SNR & 0.86$\pm$0.67   & 31 \\ %Map 21
  \NS27 &                   &              & SNR & 0.64$\pm$0.48   & 66 \\ %Map 51
  \NS28 & \ATCA{5533}{4314} & SNR/HII      & SNR & 0.45$\pm$0.15   & 63 \\ %Map 26

  \hline
  \multicolumn{6}{c}{\tiny Other Objects (\HII\ Regions?)} \\
  \hline

        & \ATCA{5438}{4144} & SNR/HII      &     & 0.17$\pm$0.07  \\ %Map 3
        & \ATCA{5438}{4240} & snr/HII      &     & 0.25$\pm$0.02  \\ %Map 4
        & \ATCA{5439}{3543} & snr$\dagger$ &     & 0.27  \\ %Map 5
        & \ATCA{5441}{3348} & bkg/snr      &     & 0.36  \\ %Map 7
        & \ATCA{5442}{4313} & SNR/HII      &     & 0.19$\pm$0.07  \\ %Map 8
        & \ATCA{5443}{4311} & SNR/HII      &     & 0.18$\pm$0.09  \\ %Map 9
        & \ATCA{5445}{3847} & SNR/HII      &     & 0.11$\pm$0.03  \\ %Map 10
        & \ATCA{5450}{4030} & SNR/HII      &     & 0.32$\pm$0.12  \\ %Map 11
        & \ATCA{5450}{3822} & SNR/HII      &     & 0.22$\pm$0.14  \\ %Map 12
        & \ATCA{5450}{4022} & SNR/HII      &     & 0.38$\pm$0.31  & 130 \\ %Map 13
        & \ATCA{5451}{3826} & SNR/HII      &     & 0.26$\pm$0.16 \\ %Map 14
        & \ATCA{5451}{3939} & SNR/HII      &     & 0.10$\pm$0.18 \\ %Map 15
        & \ATCA{5500}{4037} & SNR/HII      &     & 0.25$\pm$0.09 \\ %Map 16
        & \ATCA{5501}{3829} & SNR          &     & 0.35$\pm$0.12 & 31 \\ %Map 17
        & \ATCA{5503}{4246} & SNR/HII      &     & 0.14$\pm$0.04 \\ %Map 18
        & \ATCA{5503}{4320} & SNR/HII      &     & 0.15$\pm$0.08 \\ %Map 19
        & \ATCA{5512}{4140} & SNR/HII      &     & 0.08$\pm$0.02 \\ %Map 20
  \NS3  &                   &              & SNR & 0.24$\pm$0.31   & 26 \\ %Map 30
  \NS10 & \ATCA{5440}{4049} & SNR          & SNR & 0.35$\pm$0.15   & 63 \\ %Map 6
  \NS11 &                   &              & SNR & 0.30$\pm$0.12   & 150 \\ %Map 36
  \NS18 &                   &              & SNR & 0.32$\pm$0.32   & 69 \\ %Map 43
  \NS21 &                   &              & SNR & 0.37$\pm$0.30   & 41 \\ %Map 46
  \NS23 &                   &              & SNR & 0.31$\pm$0.08   & 43 \\ %Map 48

  \hline
  \multicolumn{6}{c}{\tiny No Signal} \\
  \hline

        & \ATCA{5423}{3648} & snr$\dagger$ \\ %Map 1
        & \ATCA{5521}{4609} & bkg/snr      \\ %Map 22
        & \ATCA{5523}{4632} & bkg/snr      \\ %Map 23
        & \ATCA{5525}{3653} & bkg/snr      \\ %Map 24
        & \ATCA{5528}{4903} & snr$\dagger$ \\ %Map 25
        & \ATCA{5541}{4033} & snr          \\ %Map 27

  \enddata
  \tablenotetext{a}{Gaussian fit returned a value too small to be deconvolved.} \\
\end{deluxetable}

\clearpage

% Table 6
\begin{deluxetable}{cccccc}
  \tabletypesize{\small}
  \tablecaption{Three neighboring sources.\label{tab:6}}
  \tablewidth{0pt}
  \tablehead{
  \colhead{1} & \colhead{2} & \colhead{3} & \colhead{4} & \colhead{5} & \colhead{6} \\
  \colhead{Source} & \colhead{\ATCA{5442}{4313}} & \colhead{\ATCA{5443}{4311}} & \colhead{\NS11} & \colhead{Carpano \#161} & \colhead{\linrat}
  }
  \startdata
  \ATCA{5442}{4313} &            & 6.57, 66.3 & 2.73, 27.6 & 20.3, 205 & 0.282$\pm$0.003 \\
  \ATCA{5443}{4311} & 6.57, 66.3 &            & 9.23, 93.2 & 25.2, 254 & 0.237$\pm$0.002 \\
  \NS11             & 2.73, 27.6 & 9.23, 93.2 &            & 18.9, 191 & 0.317$\pm$0.003 \\
  \enddata
  \tablecomments{The distance between each pair of sources is given in arcseconds and then in parsecs.}
\end{deluxetable}

%\clearpage

% Table 7
\begin{deluxetable}{ccp{11cm}}
  \tabletypesize{\small}
  \tablecaption{Sources placed in the Venn diagram of Figure \ref{fig:5}. \label{tab:7}}
  \tablewidth{0pt}
  \tablehead{
  \colhead{Venn region} & \colhead{Source count} & \colhead{Designations}}
  \startdata
  A & 17 & \NS1, S4, S5, S7 -- S9, S12 -- S17, S20, S22, S24, S25, S27 \\
  B & 3 & \ATCA{5450}{4030}, \ATCA{5450}{4022}, \ATCA{5501}{3829} \\
  C & 11 & \ATCA{5438}{4144}, \ATCA{5442}{4313} (N300-S11?), \ATCA{5443}{4311}, \ATCA{5445}{3847}, \ATCA{5450}{3822}, \ATCA{5451}{3826}, \ATCA{5451}{3939}, \ATCA{5500}{4037}, \ATCA{5503}{4246}, \ATCA{5503}{4320}, \ATCA{5512}{4140} \\
  D & 2 & \NS2 (C79), \NS19 (C123) \\
  E & 3 & \NS6 (\ATCA{5431}{3825}, C69), \NS26 (\ATCA{5515}{4439}, C34), \NS28 (\ATCA{5533}{4314}, C151) \\
  F & 3 & C60 (SNR5-R4), C72 (SNR3-R3), \NS10 (\ATCA{5440}{4049}, C12) \\
  G & 1 & C20 (SNR15-X11) \\
  \enddata
  \tablecomments{N300-S (or just S) refers to BL97 optically selected SNRs. SNR refers to P00 radio (-R) and X-ray (-X) SNRs. J refers to P04 (ATCA) radio SNRs. C refers to \citet{Carpano2005} X-ray SNRs.}
\end{deluxetable}

\clearpage

\end{document}